\documentclass{IEEEtran}

\usepackage{url}
\usepackage{subfigure}
\usepackage{graphicx}

\newcommand{\comments}[1]{}

\begin{document}

\title{Realizing ICN in 3GPP's 5G NextGen Core Architecture}

\comments{
\numberofauthors{1}
\author{
\alignauthor
Ravishankar Ravindran, Asit Chakraborti, Syed Obaid Amin, Aytac Azgin and Guoqiang Wang\\
\smallskip
\affaddr{Huawei Research Center, Santa Clara, CA, USA.}\\
\email{\large \{ravi.ravindran, asit.chakraborti, obaid.amin, aytac.azgin, gq.wang\}@huawei.com}\\
}
}

\author{Ravishankar~Ravindran,~\IEEEmembership{Senior Member,~IEEE, }
        Asit~Chakraborti,
        Syed~Obaid~Amin,
        Aytac~Azgin,~\IEEEmembership{Member,~IEEE, }
        and~Guo-Qiang~Wang
\thanks{The authors are with Huawei Research Center, Santa Clara, CA, 95050 USA e-mail(s): \{ravi.ravindran, asit.chakraborti, obaid.amin, aytac.azgin, gq.wang\}@huawei.com.}}

\maketitle


\maketitle

\begin{abstract}

The proposed 3GPP's 5G Next-generation (NextGen) Core architecture (5GC) enables the ability to introduce new user and control plane functions within the context of network slicing to allow greater flexibility in handling of heterogeneous devices and applications. In this paper, we discuss the integration of such architecture with future networking technologies by focusing on the information centric networking (ICN) technology. For that purpose, we first provide a short description of the proposed 5GC, which is followed by a discussion on the extensions to 5GC's control and user planes to support Protocol Data Unit (PDU) sessions from ICN. To illustrate the value of enabling ICN within 5GC, we focus on two important network services that can be enabled by ICN data networks. The first case targets mobile edge computing for a connected car use case, whereas the second case targets seamless mobility support for ICN sessions. We present these discussions in consideration with the procedures proposed by 3GPP's 23.501 and 23.502 technical specifications.  
\end{abstract}

\section{Introduction} \label{sec:introduction}
In contrast to 4G, 5G demands the introduction of a mix of heterogenous services. The goal of this document is to investigate and offer insights on how information-centric networking (ICN) can be enabled in the proposed 5G Next-generation Core network architecture (5GC) by leveraging its flexibility, and satisfy the needs of these new services. The reference architectural discussions in three core 3GPP specifications (\emph{i.e.}, 3GPP 23.501~\cite{23.501}, 23.502~\cite{23.502}) form the basis of our discussions.

In 5G, network slicing (NS) \cite{ngmn} is proposed to support multiple virtual networks with diverse service requirements sharing the same infrastructure. This architectural construct allows the introduction of new architectures as well, which include ICN. In \cite{5gicn}, we offered a detailed discussion on how to enable ICN (within the context of content-centric networking ~\cite{ccn}) within a network slicing framework to support an application-driven end-to-end ICN architecture, and benefits for doing it, and its ability to co-exist with other IP service slices. ICN allows a name based networking approach, which offers various network services as part of the infrastructure. These services include naming and name-based forwarding, content-based security integrated within packets, in-network caching and computing, multicasting, and multi-homing. Furthermore, ICN is also suitable for IoT applications including the ad hoc scenarios~\cite{icniot}. Although enabling ICN in 5G requires introducing new networking and service stacks, the benefits can easily surpass the cost considering the features brought in by ICN. These features include enabling a flatter network architecture, avoiding unnecessary control and user plane functions to support mobility and multicasting, and scaling content distribution without application specific overlay mechanisms, which is the norm today, such as the use of CDN frameworks.

To address the low latency and high bandwidth needs for applications, multi-access edge computing (MEC) has been proposed~\cite{etsi}, which can also be realized within 5GC through the use of local area data networking (LADN) feature. In comparison to enabling MEC using IP based networking, ICN offers three main advantages that make a compelling case for its use within 5GC:
\begin{itemize}
\item \emph{Edge Computing}: MEC aids several latency sensitive applications such as augmented and virtual reality (AR/VR) and ultra reliable and low latency class (URLLC) of applications such as autonomous vehicles. Enabling edge computing over an IP converged 5GC comes with the challenge of application level reconfiguration required to re-initialize a session whenever it is being served by a non-optimal service instance. In contrast, named-based networking, as considered by ICN, naturally supports service-centric networking, which minimizes network related configuration for applications and allows fast resolution for named service instances.
\item \emph{Edge Storage and Caching}:  The principal entity for ICN is the secured content (or named-data) object, which allows location independent data replication at strategic points in the network, or data dissemination through ICN routers by means of opportunistic caching. These features benefit both realtime and non-realtime applications, where a set of users share the same content, thereby advantageous to both high-bandwidth/low-latency applications such as AR/VR or low bandwidth IoT applications.
\item \emph{Session Mobility}: Existing long-term evolution (LTE) deployments handle session mobility using IP anchor point functions, and these typically serve a large geographical area. This design is inefficient when service instances are replicated close to radio access network (RAN) instances. Employing anchor based mobility approach in this situation shall incur high control and user plane overhead. In contrast, application bound identifier and location split principle considered for the ICN is shown to handle host mobility quite efficiently ~\cite{icnmob}.

\end{itemize}
In this paper, we carefully analyze these advantages using practical use cases over the proposed control and user plane extensions of the 5GC architecture to formally support ICN sessions.

\comments{
Enabling ICN in 5G offers multiple advantages with respect to edge computing. This is one of the features 5GC enables through local service point-of-presence using compute and storage resource physically close the base station to. This enables distributed service logic execution unlike the central resource setup located behind the evolved packet core (EPC) in LTE today, adding significantly to end-to-end latency. We discuss ICN based edge computing along with session and content distribution scenarios as use cases to justify the benefits of enabling ICN in the end points and part of the 5GC architectural extensions to support ICN networking.
}

The remaining sections are laid out as follows. In {Section~\ref{sec:icn}}, we provide a brief introduction to ICN. We discuss the design principles for 5GC framework that allows the introduction of new networking architectures in {Section~\ref{sec:design}}. In {Section~\ref{sec:5gnextgen}}, we present a summary for the 5GC proposal, with focus on control/user plane functions that are relevant to supporting ICN. Next, in {Section~\ref{sec:5gcicn}}, we  discuss an ICN enabled 5GC with normative interface extensions and control/user plane functions relevant to supporting ICN-enabled PDU sessions. 
In {Section~\ref{sec:usecase}}, we offer discussions on two services enabled through formal ICN support: first being, MEC, considering a connected car scenario; second, seamless session mobility, considering an ICN enabled 5GC. We present our final remarks in Section~\ref{sec:conclude}. 

\section{Information-centric Networking}\label{sec:icn}
Information-centric Networking (ICN)~\cite{icnsurvey} is a result of a global research effort on future network architectures and enables features such as: \begin{itemize}
\item Name based networking of resources corresponding to contents, services, devices and network domains;
\item Session-less transport through per-hop name resolution (of the requested resource), which also enables 5G-targeted features such as mobility, multicasting, and multi-homing;
\item Exploiting  network embedded compute/storage resources that are virtualizable among heterogenous services;
\item Network layer security to authenticate user requests and content objects and to allow location-independent caching and computing, which is a desired feature for ICN applications and infrastructure providers; and
\item Suitability to both adhoc- and infrastructure-based IoT environments, where the information-centric nature of IoT applications matches with what the ICN infrastructure provides.
\end{itemize}

Though ICN has been an active area of research, many research challenges still remain~\cite{rfc-icn-challenge} in various aspects of the architecture and with regards to business feasibility.
\comments{
We next summarize 3GPP's 5GC proposal and followed by extensions required to formally support ICN PDU sessions.
}

\section{5GC Design Principles}\label{sec:design}
The design for 5GC architecture is based on the following design principles that allow it to support new service networks like ICN efficiently compared to LTE networks:
\begin{itemize}

\item \emph{Control and User plane split $($CUPS$)$}: This design principle moves away from LTE's vertically integrated control/user plane design, \emph{e.g.}, Serving Gateway (S-GW) and Packet Data Network Gateway (P-GW), to one espousing a network function virtualization (NFV) framework with modular network functions separated from the hardware for service-centricity, flexibility and programmability. In doing so, network functions can be implemented both physically and virtually, while allowing each to be customized and scaled based on their individual requirements, also allowing the realization of multi-slice co-existence. This also allows the introduction of user plane functions (UPF)\footnote{UPF is the generalized logical data plane function corresponding to a UE PDU session. UPFs can play many roles, such as, being an uplink flow classifier (UL-CL), a PDU session anchor point, a branching point function, or one based on new network architectures like ICN.} with new control functions, or re-using/extending the existing ones, to manage the new user plane realizations.

\item \emph{Decoupling of RAT and Core Network} : Unlike LTE's unified control plane for access and the core, 5GC offers control plane separation of RAN from the core network. This allows the introduction of new radio access technologies (RAT) and map multiple heterogenous RAN sessions to arbitrary core network slices based on service requirements.

\item \emph{Non-IP PDU Session Support}: A PDU session is defined as the logical connection between the user equipment (UE) and the data network (DN). 5GC offers a scope to support both IP and non-IP PDU (termed as "unstructured" payload), and this feature can potentially allow the support for ICN PDUs by extending or re-using the existing control functions.

\item \emph{Service Centric Design}: 5GC service orchestration and control functions, such as naming, addressing, registration/authentication and mobility, will utilize cloud based service APIs. Doing so enables opening up interfaces for authorized service function interaction and creating service level extensions to support new network architectures. These APIs include the well accepted Get/Response and Pub/Sub approaches, while not precluding the use of procedural approach between functional units (where necessary).

\end{itemize}

\begin{figure}[ht]
\centering
\includegraphics[scale=0.4,clip,angle=270]{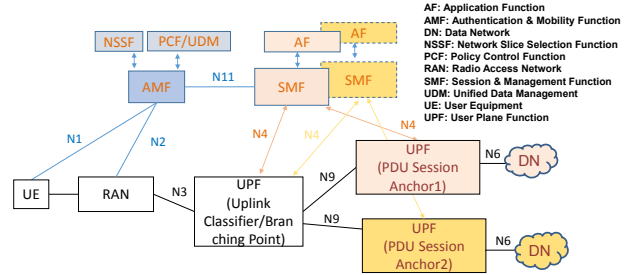}
\caption{5G Next Generation Core Architecture.}
\label{fig:5gnextgen}
\vspace{-.1in}
\end{figure}

\section{5G NextGen Core Architecture}  \label{sec:5gnextgen}
In this section, for brevity purposes, we restrict the discussions to the control and user plane functions relevant to an ICN deployment. More exhaustive discussions on the various architecture functions, such as registration, connection and subscription management, can be found in~\cite{23.501} and~\cite{23.502}.

In Figure~\ref{fig:5gnextgen}, we show one variant of a 5GC architecture from \cite{23.501}, for which the functions of UPF's branching point and PDU session anchoring are used to support inter-connection between a UE and the related service or data networks (or DNs). In 5GC, control plane functions can be categorized as: ($i$) common control plane functions that are common to all slices and which include the Authentication and Mobility Function (AMF),  Network Slice Selection Function (NSSF), Policy Control Function (PCF), and Unified Data Management (UDM) among others, and ($ii$) shared or slice specific control functions, which include the Session and Management Function (SMF) and the Application Function (AF).

Among these functions, AMF serves multiple purposes: 1) device authentication and authorization; 2) security and integrity protection to non-access stratum (NAS) signaling; 3) tracking UE registration in the operator's network and mobility management functions as the UE moves among different RANs, each of which might be using different radio access technologies (RATs).

NSSF is used to handle the selection of a particular slice for the PDU session request by the UE using the Network Slice Selection Assistance Information (NSSAI) parameters provided by the UE \cite{ns} and the configured user subscription policies in PCF and UDM functions. Compared to LTE's evolved packet core (EPC), where PDU session states in RAN and core are synchronized with respect to session management, 5GC decouples this using NSSF by allowing PDU sessions to be defined prior to a PDU session request by a UE (for other differences see \cite{lteversus5g}). This de-coupling allows dynamic and policy based inter-connection of UE flows with slices provisioned in the core network.

SMF is used to handle IP anchor point selection and addressing functionality, management of the user plane state in the UPFs (such as in uplink classifier (UL-CL) and branching point functions) during PDU session establishment, modification and termination, and interaction with RAN to allow PDU session forwarding in uplink/downlink (UL/DL) to the respective DNs.

In the data plane, UE's PDUs are tunneled to the RAN using the 5G RAN protocol~\cite{5gnr}. From the RAN, the PDU's five tuple header information (IP source/destination, port, protocol etc.) is used to map the flow to an appropriate tunnel from RAN to UPF. 

\section{5G NextGen Core Architecture with ICN Support}  \label{sec:5gcicn}

In this section, we focus on control and user plane enhancements required to enable ICN within 5GC, and identify the interfaces that require extensions to support ICN PDU sessions. Explicit support for ICN PDU sessions within access and 5GC networks will enable applications to leverage the core ICN features, while offering it as a service to 5G users.

For an ICN-enabled 5GC network, the assumption is that the UE may have applications that can run over ICN or IP, with, for instance, UE's operating system offering applications to operate over ICN-~\cite{ccnx} or IP-based networking sockets with appropriate transport convergence layer to identify session flows and mux/de-mux flows in the UL and DL from the 5G New Radio (5G-NR) layer~\cite{5gnr}. More insights on co-existence of ICN and IP in UE has been presented in ~\cite{prakash}. There may also be cases where UE is exclusively based on ICN. In either case, we identify an ICN enabled UE as ICN-UE. This discussion can be seen in light of one or many ICN network instances operating as a slice in an architecture driven using a network slicing (NS) framework~\cite{5gicn}. This discussion also borrows ideas from~\cite{23.799}, which offers a wide range of architectural discussions and proposals on enabling slices and managing multiple PDU sessions with local networks (enabling MEC) and its associated architectural support (in the service, control and data planes) and procedures within the context of 5GC.

Figure~\ref{fig:5gcicn} shows the proposed ICN-enabled 5GC architecture. In the figure, new/modified functional components identify an ICN user and control plane required to interconnect an ICN-DN with 5GC. The interfaces and functions that require extensions to enable ICN as a service in 5GC are identified in the figure with a '\emph{++}' symbol.

We next summarize the control, user plane and normative interface extensions that formally help with ICN support.

\comments{
\begin{figure}[ht]
\centering
\includegraphics[scale=0.4,clip,angle=270]{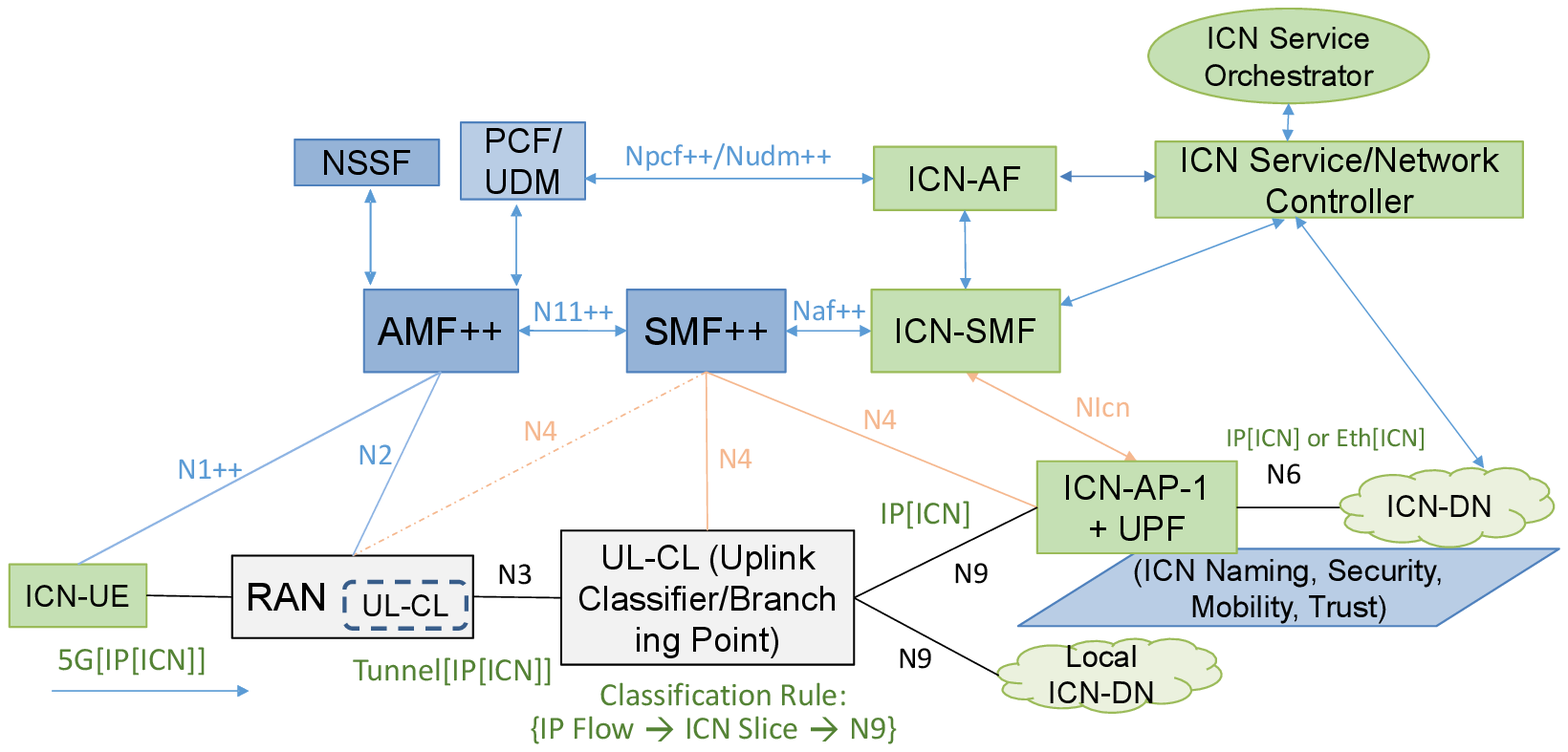}
\caption{5G Next Generation Core Architecture with ICN Support.}
\label{fig:5gcicn}
\end{figure}

}

\begin{figure}[ht]
\centering
\includegraphics[scale=0.5, clip]{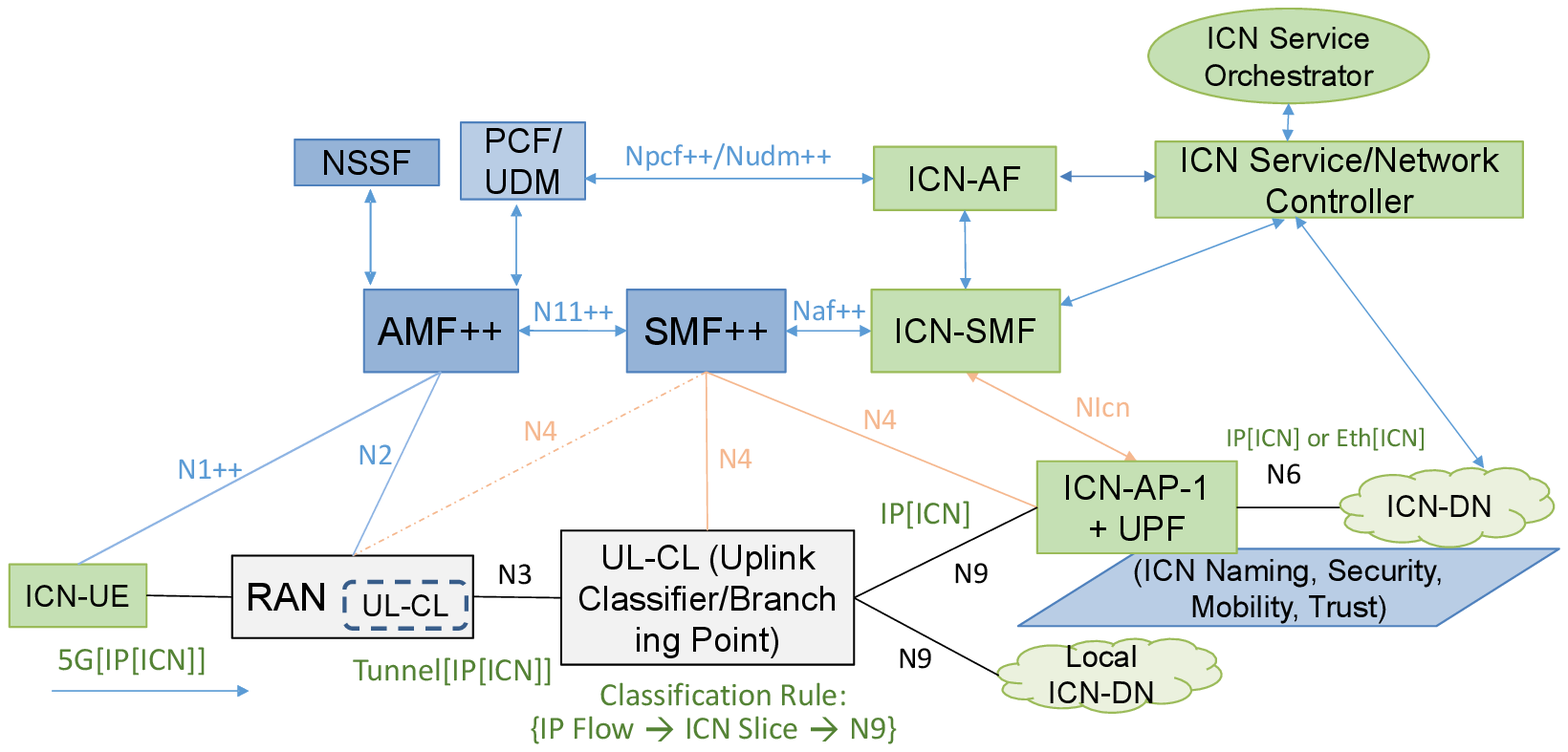}
\caption{5G Next Generation Core Architecture with ICN Support.}
\label{fig:5gcicn}
\end{figure}

\subsection{Control Plane Extensions}  \label{sec:5gcicncontrolplane}
To support interconnection between ICN UEs and the appropriate ICN DN instances, we propose the following control plane extensions.

\subsubsection{Authentication and Mobility Function (AMF++)}
Applications in the UEs have to be authorized to access ICN DNs. For this purpose, as in~\cite{23.501}, operator enables ICN as a DN offering ICN services. As a network service, ICN-UE should also be subscribed to it and this is imposed using the PCF and unified data management (UDM) functions, which may interface with the ICN Application Function (ICN-AF) for policy management of ICN PDU sessions. Hence, if the UE policy profile in the UDM doesn't enable this feature, then the ICN applications in the UE will not be allowed to connect to ICN DNs.

To enable ICN stack in the UE, AMF++ function has to be enabled with the capability of authenticating UE's ICN applications. AMF++ can potentially be extended to also support ICN specific bootstrapping and forwarding functions (such as naming and security) to configure UE's ICN layer and the applications, and the forwarding rules for the name prefixes that bind the flows to appropriate 5G-NR logical tunnel or slice interfaces. These functions can also be handled by the ICN-AF and ICN control function in the UE after setting PDU session state in 5GC. Furthermore, during handover, AMF++ extensions will support ICN related state information signalling to the ICN UPFs to support seamless session handover handling in the ICN-DN.

\subsubsection{Session Management Function (SMF++)}
Once a UE is authenticated to access ICN service in network, SMF manages to connect UE's ICN PDU sessions to the ICN DN in the UL/DL. For this, SMF++ creates appropriate PDU session policies in the UPF, which include UL-CL and ICN anchor point (ICN-AP) (discussed in Section~\ref{sec:5gcicndataplane}) through the ICN-SMF. 

SMF++ interfaces with AMF++ to enable ICN specific user plane state, which includes IP address configuration and associated traffic filter policy to inter-connect the RAN (UE and the base station (BS)) with the appropriate core network slice. Furthermore, AMF++ sets appropriate state in the RAN that directs ICN flows to chosen ICN UL-CL.

\subsubsection{ICN Session Management Function (ICN-SMF)}
ICN-SMF serves as control plane for the ICN state managed in ICN-AP. This function interacts with SMF++ to obtain and also push ICN PDU session management information for the creation, modification and deletion of ICN PDU sessions in ICN-AP. For instance, when new ICN slices are provisioned by the ICN service orchestrator, ICN-SMF requests a new PDU session to the SMF that extends to the RAN. While SMF manages the tunnels to interconnect ICN-AP to UL-CL, ICN-SMF creates the appropriate forwarding state in ICN (using the forwarding information base or FIB) to enable ICN flows over appropriate tunnel interfaces managed by the SMF. In addition, it also signals resource management rules to share compute, bandwidth, storage/cache resources among multiple slice instances co-located in the ICN-AP.

\subsubsection{ICN Application Function (ICN-AF)}
ICN-AF represents the application controller function that interfaces with ICN-SMF in the 5GC to manage the ICN service function and network state in ICN-AP. In addition, ICN-AF also interfaces with PCF and UDM functions to transfer user profile, subscription and authentication policies, and slice details required to validate ICN-UE's PDU session request and map to appropriate ICN slice through NSSF. ICN-AF interfaces with ICN service and network controller functions, which can influence both ICN-SMF and SMF to steer ICN PDU session traffic and serve edge service functions.

\subsubsection{Normative Interface Extensions}

\begin{itemize}

\item \emph{N1++, N11++}: This extension enables ICN specific control functions to support ICN authentication, configuration and programmability of an ICN-UE via AMF++ and SMF++, and also impose QoS requirements, handle mobility management of an ICN PDU session in 5GC based on service requirements.

\item \emph{N4}: Though this signaling is service agnostic, as discussed in Section~\ref{sec:5gcicndataplane}, future extensions may include signaling to enable ICN user plane features in these network functions. The extension of N4 to RAN is to handle the case when UPF function collocates with the RAN instance to enable localized ICN DNs.

\item \emph{NIcn}: This extension serves two functions: (\emph{i}) control plane programmability using name based forwarding rules to manage edge service functions and ICN PDU sessions applicable to 5GC in ICN-AP; (\emph{ii}) control plane extensions to enable ICN mobility anchoring at ICN-AP, in which case it also acts as a point of attachment (POA) for ICN flows. Features such as ICN mobility as a service can be supported with this extension as further discussed in~\cite{5gicn}.

\item\emph{Naf++}: This extension supports 5GC control functions such as  naming, addressing, mobility, and tunnel management for ICN PDU sessions to interact with SMF++ and AMF++.  

\item\emph{Npcf++,Nudm++}:  This extension creates an interface to push ICN PDU session requirements to PCF and UDM functions that interact with the ICN-AF function for ICN slice specific configuration. These requirements are enforced at various steps, for instance, during ICN application registration, authentication, slice mapping, and provisioning of resources for these PDU sessions in the UPF.

\end{itemize}

\subsection{User Plane Extensions}  \label{sec:5gcicndataplane}
As explained in detail in \cite{23.501}, UPFs are service agnostic functions, hence extensions are not required as such to operate an ICN-DN. The interconnection of a UE to an ICN-DN comprises of two segments, one from RAN to UL-CL and the other from UL-CL to ICN-AP. These segments use IP tunnelling constructs, where the service semantic check at UL-CL is performed using IP's five tuples to determine both UL and DL tunnel mappings. We summarize the relevant UPFs and the interfaces for handling ICN PDU sessions as follows.

\subsubsection{Uplink Classifer (UL-CL)}
UL-CL enables classification of flows based on source or destination IP address and steers the traffic to an appropriate network or service function anchor point. If the ICN-AP is identified based on service IP address associated with the ICN-UE's flows, UL-CL checks the source or destination address to direct traffic to an appropriate ICN-AP. As UL-CL is a logical function, it can also reside in RAN, as shown in Figure~\ref{fig:5gcicn}, where traffic classification rules can be applied over 5G-NR protocols to determine the tunnel to forward the ICN payload towards the next ICN-AP. For native ICN UE, ICN shall be deployed on Layer-2 MAC, hence there may not be any IP association; for such packet flow, new classification schema shall be required. 

\subsubsection{ICN Anchor Point (ICN-AP)}
ICN-AP is where the 5GC PDU sessions terminate and ICN service network begins. Compared to the traditional anchor points as in P-GW, the ICN-AP is also a service gateway as it can host services or cache content enabled through the ICN architecture. The ICN-AP also includes the UPF functions to manage multiple tunnel interfaces enabling the relay of ICN PDU flows to appropriate UL-CL instances in the DL. Note that there may be multiple ICN-APs serving different ICN services or slices. ICN-AP also manages other ICN functions such as enforcing the dynamic name based forwarding state, mobility state, in-network service function management, resource management with respect to sharing caching, storage, and compute resources among multiple services~\cite{5gicn}.

\subsubsection{ICN Data Network (ICN-DN)}
ICN-DN represents a set of ICN nodes used for ICN networking and with heterogeneous service resources such as storage and computing points. 

\subsubsection{User Plane Interface Extensions}
\begin{itemize}
\item \emph{N3}: Though the current architecture supports heterogeneous service
      PDU handling, future extensions can include user plane interface
      extensions to offer explicit support to ICN PDU session traffic,
      for instance, an incremental caching and computing function in RAN
      or UL-CL to aid with content distribution.

\item \emph{N9}: Extensions to this interface can consider UPFs to enable
      richer service functions, for instance to aid context processing.
      In addition extensions to enable ICN specific encapsulation to
      piggyback ICN specific attributes such as traffic characteristics
      between the UPF branching point and the ICN-AP.  The intermediate
      nodes between the UL-CL and the ICN-AP can also be other caching
      points.

\item \emph{N6}: This interface is established between the ICN-AP and the ICN-DN, whose networking elements in this segment can be deployed as an overlay or as a native Layer-3 network.
\end{itemize}

\section{Use Case Scenarios} \label{sec:usecase}
Based on the 5G-ICN architecture presented in the previous section, we next present two useful network services that can be enabled using ICN-DNs. The first case targets ICN-MEC considering a connected car scenario highlighting the use of ICN's edge compute, storage/caching features and the other ICN session mobility that is handled by the ICN-AP with minimal support from 5GC.

\begin{figure}[t]
\centering
\includegraphics[scale=0.45,clip,angle=270]{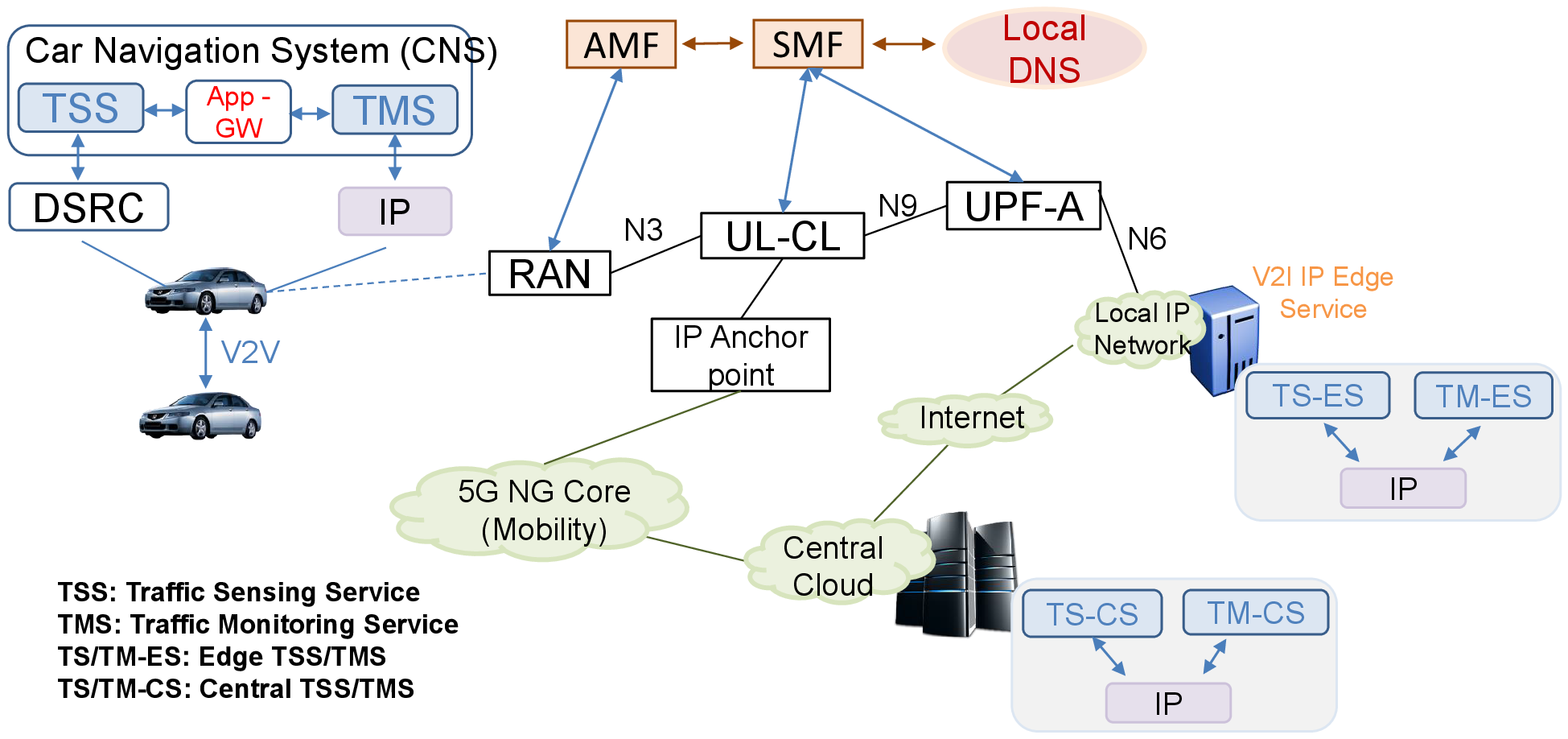}
\caption{Traffic sensing and monitoring application over IP-MEC.}
\label{fig:ipmec}
\vspace{-.1in}
\end{figure}

\begin{figure}[t]
\centering
\includegraphics[scale=0.45]{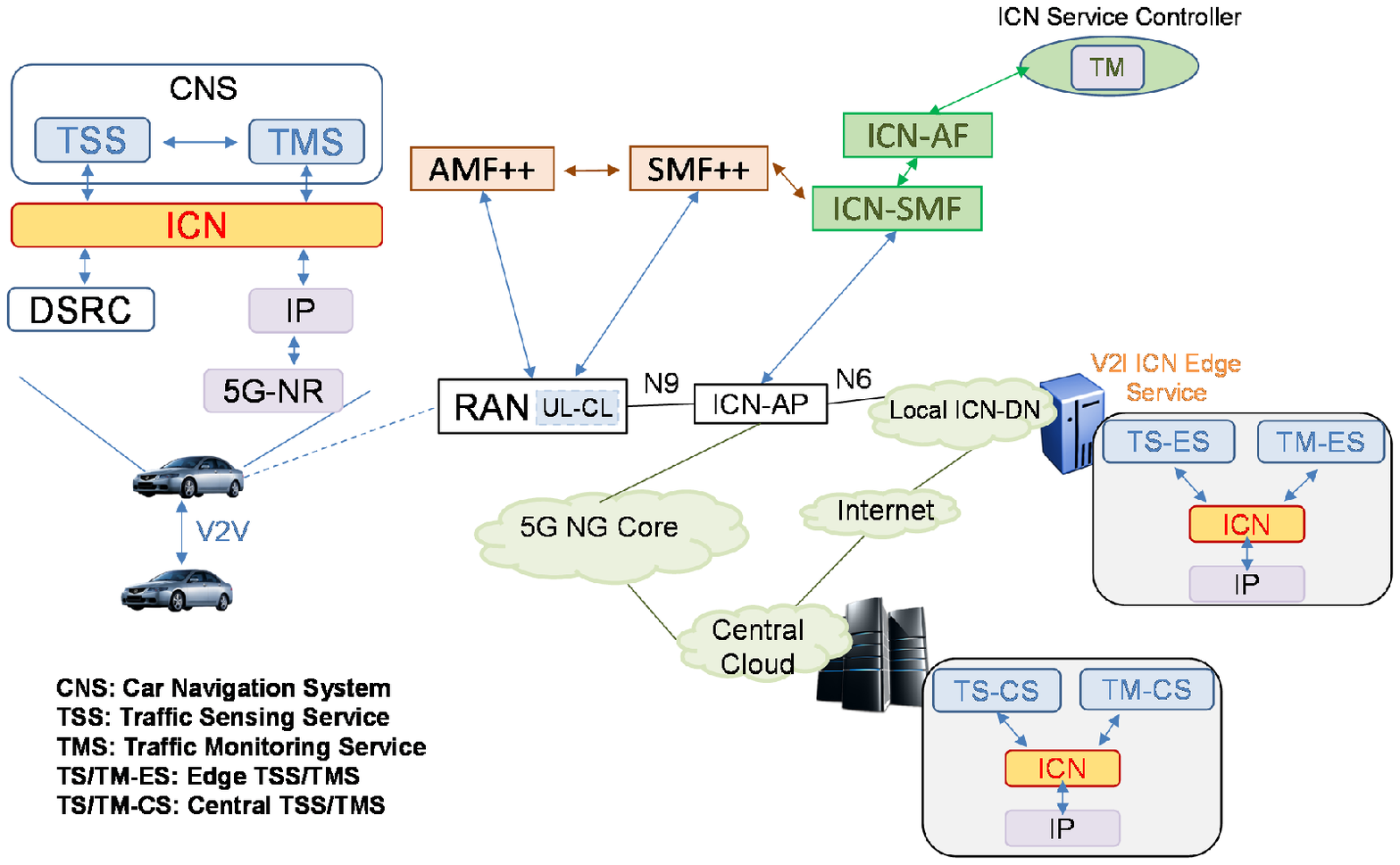}
\caption{Traffic sensing and monitoring application over ICN-MEC.}
\label{fig:icnmec}
\vspace{-.1in}
\end{figure}

\subsection{Mobile Edge Computing}
We consider here a connected vehicle scenario, where the car's navigation system (CNS) uses data from the edge traffic monitoring (TM-E) service instance  to offer rich insights on the road conditions (such as real-time congestion assisted with media feeds). This is aided using traffic sensing (TS) information collected through vehicle-to-vehicle (V2V) communication over dedicated short-range communications (DSRC) radio~\cite{dsrc} by the TS-E, or using road-side sensor units (RSU) from which this information can be obtained. The TS-E instances then push this information to a central traffic sensing instance (TS-C). This information is used by the central traffic monitoring service (TM-C) to generate useable navigation information, which can then be periodically pushed to or pulled by the edge traffic monitoring service (TM-E) to respond to requests from vehicle's CNS.

For this scenario, our objective is to compare advantages of offering this service over an IP based MEC versus one based on ICN. We can generalize the following discussion to other MEC applications as well. The realization of these two scenarios within ICN-enabled 5GC is shown in Figures~\ref{fig:ipmec} and \ref{fig:icnmec}.

\subsubsection{IP-based MEC}
When a vehicle's networking system comes online, it first undergoes an attachment process with the 5G-RAN, which includes authentication, IP address assignment and DNS discovery. The attachment process is followed by PDU session establishment, which is managed by SMF signaling to UL-CL and the UPF instance. When the CNS application initializes, it assumes this IP address as its own ID and tries to discover the closest service instance. Local DNS then resolves the service name to a local MEC service instance. Accordingly, CNS learns the IP service point address and uses that to coordinate between traffic sensing and monitoring applications.

Following are the challenges with this design: (\emph{i}) at the CNS level, non-standardization of the naming schema results in introducing an application level gateway to adapt the sensing data obtained from DSRC system to IP networks, which becomes mandatory if the applications are from different vendors; (\emph{ii}) as the mobility results in handover between RAN instances, service-level or 5GC networking-level mechanisms need to be initiated to discover a better TM-E instance, which may affect the service continuity and result in session reestablishment that introduces additional control/user plane overheads; (\emph{iii}) considering data confidentiality needs, authentication and privacy control are offered through an SSL/TLS mechanism over the transport channel, which has to be re-established whenever the network layer attributes are reset.

\subsubsection{ICN-based MEC}
If the CNS application is developed over ICN, ICN allows the same named data logic to operate over heterogenous interfaces (such as DSRC radio and  IP-over-5G link), thereby avoiding the need for application layer adaptation.\footnote{Here, adaptation is resolved by the named-networking architecture to allows data to be seamlessly exchanged between heterogenous application functions.}

We can list the advantages of using ICN-based MEC as follows: (\emph{i}) as vehicles within a single road segment are likely to seek the same data, ICN-based MEC allows to leverage opportunistic caching and storage enabled at ICN-AP, thereby avoiding service level unicast transmissions; (\emph{ii}) processed and stored traffic data can be easily contextualized to different user requirements; (\emph{iii}) appropriate mobility handling functions can be used depending on mobility type (as consumer or producer) \footnote{Specifically, when an ICN-UE moves from one RAN instance to another, the next IP hop, which identifies the ICN-AP function, has to be re-discovered. Unlike the IP-MEC scenario, this association is not exposed to the applications. As discussed in Section~\ref{sec:5gcicncontrolplane}, control plane extensions to AMF and SMF can enable re-programmbility of the ICN layer in the vehicle to direct it towards a new ICN-AP, or to remain with the same ICN-AP, based on optimization requirements.};(\emph{iv}) as ICN offers content-based security, produced content can be consumed while authenticating it at the same time (\emph{i.e.}, allowing any data produced to diffuse to its point of use through named data networking).

\begin{figure}[t]
\centering
\includegraphics[scale=0.3,clip,angle=270]{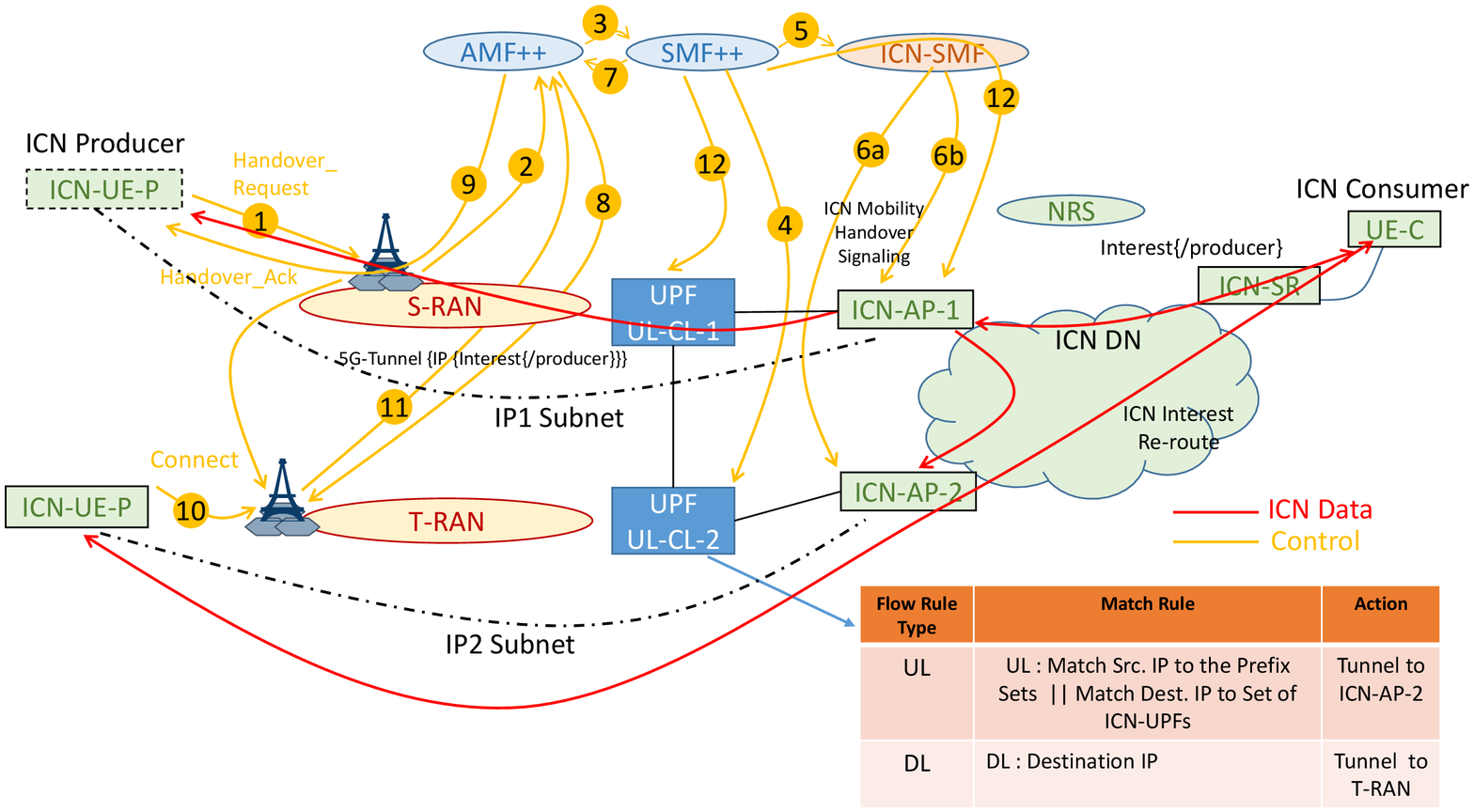}
\caption{ICN Seamless Mobility in 5GC.}
\label{fig:icnmobility}
\vspace{-.1in}
\end{figure}

\subsection{ICN Session Mobility}
Mobility scenario, which is shown in Figure~\ref{fig:icnmobility}, assumes a general ICN-UE handover from S-RAN to T-RAN, where each of them is served by different UPFs, \emph{i.e.}, UL-CL-1 and UL-CL-2. We also assume that UL-CL-1 and UL-CL-2 use different ICN-APs as gateways, referred to as ICN-AP-1 and ICN-AP-2. From an ICN perspective, we discuss here the producer mobility case, which can be handled in multiple ways, one of which is proposed in~\cite{icnmob}.\footnote{Note that, the details of the ICN mobility solution provided in \cite{icnmob} are orthogonal to this discussion.} In the figure, ICN-UE refers to an application producer (\emph{e.g.}, video conferencing application), from which ICN consumers request real-time content. Here we also assume the absence of any direct physical interface, \emph{Xn}, between the two RANs. The current scenario follows the handover procedures discussed in~\cite{23.502}, following a \emph{make-before-break} approach, with focus here on integrating it with an ICN-AP and ICN-DN, where mobility state of the ICN sessions are handled.\footnote{Note that the difference from an anchor based mobility solution employed in LTE or that proposed by 5G is same as the argument presented in the MEC scenario.}. The overall signalling overhead depends on the deployment models discussed in Section~\ref{sec:5gcicn}. Here we consider the case when RAN, UL-CL and ICN-AP are physically disjoint; however in the case when RAN and UL-CL are co-located then a part of the signalling to manage the tunnel state is localized, which then improves the overall overhead efficiency. This can be further extended to the case of co-locating ICN-APs along with the RAN and UL-CL, leading to further simplification of the mobility signalling. 

Next, we discuss the high-level steps involved during handover, which are also enumerated in Figure~\ref{fig:icnmobility}.

\textbf{Step 1}: When the ICN-UE decides to handover from S-RAN to T-RAN, ICN-UE signals the S-RAN with a \emph{handover-request} indicating the new T-RAN it is willing to connect. This message includes the affected PDU session IDs from the 5GC perspective, along with the ICN names that require mobility support.

\textbf{Step 2}: S-RAN then signals the AMF serving the ICN-UE about the handover request. The request includes the T-RAN details, along with the affected ICN PDU sessions.

\textbf{Step 3}: Here, when SMF receives the ICN-UE's and the T-RAN information, it identifies UL-CL-2 as the better candidate to handle the ICN PDU sessions to T-RAN. In addition, it also identifies ICN-AP-2 as the appropriate gateway for the affected ICN PDU sessions.

\textbf{Step 4}: SMF signals the details of the affected PDU sessions along with the traffic filter rules to switch the UL traffic from UL-CL-2 to ICN-AP-2 and DL flows from UL-CL-2 to T-RAN.

\textbf{Step 5}: SMF then signals ICN-SMF about the PDU session mobility change along with the information on UL-CL-2 for it to provision the tunnel between ICN-AP-2 and UL-CL-2.

\textbf{Step 6}: Based on the signaling received on the ICN PDU session, ICN-SMF identifies the affected gateways, \emph{i.e.}, ICN-AP-1 and ICN-AP-2: (\emph{i}) ICN-SMF signals ICN-AP-2 about the affected PDU session information to update its DL tunnel information to UL-CL-2. Then, based on the ICN mobility solution, appropriate ICN mobility state to switch the future incoming Interests from ICN-AP-1 to UL-CL-2; (\emph{ii}) ICN-SMF also signals ICN-AP-1 with the new forwarding label~\cite{icnmob}  to forward the incoming Interest traffic to ICN-AP-2. This immediately causes the new Interest payload for the ICN-UE to be send to the new ICN gateway in a proactive manner.

\textbf{Step 7}: ICN-SMF then acknowledges SMF about the successful mobility update. Upon this, the SMF then acknowledges AMF about the state changes related to mobility request along with the tunnel information that is required to inter-connect T-RAN with UL-CL-2.

\textbf{Step 8}: AMF then updates the T-RAN PDU session state in order to tunnel ICN-UE's PDU sessions from T-RAN to UL-CL-2. This is followed by initiating the RAN resource management functions to reserve appropriate resources to handle the new PDU session traffic from the ICN-UE.

\textbf{Step 9}: AMF then signals the \emph{handover-ack} message to the UE, signaling it to handover to the T-RAN.

\textbf{Step 10}: UE then issues a \emph{handover-confirm} message to T-RAN. At this point, all the states along the new path comprising the T-RAN, UL-CL-2 and ICN-AP-2 is set to handle UL-DL traffic between the ICN-UE and the ICN-DN.

\textbf{Step 11}: T-RAN then signals the AMF on its successful connection to the ICN-UE. AMF then signals S-RAN to remove the allocated resources to the PDU session from the RAN and the tunnel state between S-RAN and UL-CL-1.

\textbf{Step 12}: AMF then signals SMF about the successful handover, upon which SMF removes the tunnel states from UL-CL-1. SMF then signals the ICN-SMF, which then removes the ICN mobility state related to the PDU session from ICN-AP-1. Also at this point, ICN-SMF can signal the ICN-NRS (directly or through ICN-AP-2) to update the UE-ID resolution information, which now points to ICN-AP-2~\cite{icnmob}.

Note that, inter-RAN handover mapping to the same UL-CL represents a special case of the above scenario.

\section{Conclusion}\label{sec:conclude}
In this paper, we explore the feasibility of realizing future networking architectures like ICN within the proposed 3GPP's 5GC architecture. Towards this, we summarized the design principles that offer 5GC the flexibility to enable new network architectures. We then discuss a 5GC architecture along with the user/control plane extensions required to handle ICN PDU sessions. We then apply the proposed architecture to two relevant services that ICN networks can enable: first, mobile edge computing over ICN versus the traditional IP approach considering a connected car scenario, and argue the architectural benefits of using an ICN-enabled design; second, handling ICN PDU session mobility in ICN-DN rather than using IP anchor points, with minimal support from 5GC.


\small
\bibliographystyle{unsrt85}
\bibliography{paper-demo-3gpp}

\end{document}